\documentclass[a4paper,10pt]{article}
\usepackage[a4paper, total={5.5in, 8in}]{geometry}
\usepackage[utf8]{inputenc}
\usepackage{graphicx}
\newcommand{\be}{\begin{equation}}
\newcommand{\ee}{\end{equation}}
\newcommand{\bea}{\begin{eqnarray}}
\newcommand{\eea}{\end{eqnarray}}
\newcommand{\ba}[1]{\begin{array}{#1}}
\newcommand{\ea}{\end{array}}
\newcommand{\nn}{\nonumber}
\newcommand{\vp}{\vec{p}}
\newcommand{\ep}{\epsilon}
\newcommand{\om}{\omega}

\newcommand{\vk}{\vec k}
\newcommand{\vq}{\vec q}
\begin{document}

%opening
\title{From Non-interacting to Interacting Picture of Thermodynamics and Transport Coefficients for Quark Gluon Plasma}

\author{Sarthak Satapathy$^\star$, Souvik Paul$^{\dagger,\star}$, Ankit Anand$^{\dagger,\star}$, 
Ranjesh Kumar$^{\dagger,\star}$,
\\
Sabyasachi Ghosh$^\star$}
\date{}
\maketitle
%\begin{document}
\begin{center}
{$^\star$Indian Institute of Technology Bhilai, GEC Campus, Sejbahar, Raipur 492015, Chhattisgarh, India}
\\
{$^\dagger$ Department of Physical Sciences,Indian Institute of Science Education and Research Kolkata, Mohanpur, West Bengal 741246, India}
\end{center}

\begin{abstract}
We have attempted to build first some simplified model to map the interaction of quarks and gluons, 
which can be contained by their thermodynamical quantity like entropy density, obtained from
calculation of lattice quantum chromo dynamics (LQCD). With respect to entropy density of the standard 
non-interacting massless quark gluon plasma (QGP), its interacting values from LQCD simulation are reduced
as we go from higher to lower temperature through the cross-over of quark-hadron phase transition. 
By parameterizing increasing degeneracy factor or increasing interaction-fugacity or decreasing
thermal width of quarks and gluons with temperature, we have matched LQCD data.Using that interaction
picture, shear viscosity and electrical conductivity are calculated. For getting nearly perfect fluid
nature of QGP, interaction might have some role when we consider temperature dependent
thermal width.
\end{abstract}

\section{Introduction}
% The work is motivated by different quasi-particle descriptions of QGP matter to describe various properties of QGP.
%  In these models the the fugacity of the system is used to formulate the EOS Ref-[Vinod Chandra's Papers] which describe various properties of RHIC. 
% The procedure involves extraction of the equilibrium distribution functions for quarks and gluons as in [REF- Vinod Chandra's papers] and form the EOS in terms of the chemical potential of the partons. Bulk 
% and transport properties of QGP. The EOS work for particular order of the coupling constant with restrictions imposed by perturbative and non-perturbative contributions. In all the EOS derived 
% the interaction effects are due to the effective chemical potential for the gluons and quarks. \\
%
% Quark Gluon plasma and it's thermodynamics have been well understood by employing the quasi particle model(QPM) Ref-[] and interpreting the LQCD equation of
% state. Quasi particle models have been constructed based on prescriptions such as mass as a function of temperature Ref-[Peshier] and introducing interactions
% through effective fugacity Ref-[Vinod Chandra]. QPM have succesfully described pure SU(3) gauge theory lattice data through the effective fugacity Ref-[Vinod Chandra, 
% Ravishankar 2009]. EOS obtained from quasi-particle model has been used to make predictions for RHIC Ref-[Vinod, Ravindra, Ravishankar-2010] and study viscosity and thermodynamic
% propeties Ref-[Vinod, Ravishankar] of QGP in RHIC. 
%
Thermodynamics of quark gluon plasma (QGP) can be well described by lattice quantum chromodynamics (LQCD)
calculation, which predicted a quark-hadron phase transition since a long time~\cite{Karsh} 
and its numerical estimations have gone through several up-gradation~\cite{Sayantan}. 
From the direction of perturbative quantum chromodynamics (QCD) theory, a re-summation method,
known as hard thermal loop (HTL) approach, is well studied historically, whose
latest status can be found in Ref.~\cite{HTL_rev}. 
In this direction it is found that three loop order calculations of
HTL perturbation theory~\cite{Anderson_HTL,Anderson_HTL2} can well describe the LQCD results of thermodynamics
beyond the transition temperature.
The finite density extension of this three loop calculations has been studied 
in Ref.~\cite{Najmul,Najmul2}. Similar to quark temperature domain, LQCD thermodynamics
of hadronic temperature domain can be well described by simple ideal hadron resonance
gas (HRG) model~\cite{HRG_rev}. To describe the LQCD thermodynamics of both quark
and hadronic temperature domain, there are very successful effective 
QCD models~\cite{NJL,NJL2,NJL3}, which can explain the smooth cross-over results of quark-hadron
transition. 
In this context, Quasi particle model (QPM) is a widely used 
framework~\cite{Gorenstein,Peshier,Heinz,Bluhm,Bannur,Salvatore,Meisinger,
Ruggieri,Chandra_PRC07,Chandra_EPJC09,Chandra_PRD11}
to describe the thermodynamics of QGP through entire temperature zone. 
Some attempts~\cite{Gorenstein,Peshier,Heinz,Bluhm} made 
by incorporating temperature dependent masses of medium constituents, which
become quite popular and standard.
Ref.~\cite{Bannur} has built a self-consistent equation, where gluon mass is considered
as plasma frequency, depending upon density of the system. 
Refs.~\cite{Chandra_PRC07,Chandra_EPJC09,Chandra_PRD11} have built interestingly a quasi-particle model
by introducing effective fugacity parameter in thermal distribution functions
of quarks and gluons. In present work, we have first followed the effective fugacity 
methodology, adopted by Refs.~\cite{Chandra_PRC07,Chandra_EPJC09,Chandra_PRD11}, and 
attempted to reproduce qualitatively an existing quasi-particle model. Then we have tried
to find more alternative methodologies, which can match same LQCD thermodynamics. By comparing
them, we have tried to extract some common qualitative message. 

% QPM has been succesfully used to describe lattice QCD(LQCD) equation of state(EOS) for pure 
% SU(3) gauge theory ~\cite{Ruggieri} by introducing interactions through effective fugacity ~\cite{Vinod1}.
% This idea has been further utilized to describe (2+1) flavor LQCD EOS with physical quark masses ~\cite{Vinod2} 
% where the mapping is found to be exact for the EOS. EOS 
% obtained from quasi-particle model has been used to make predictions for RHIC ~\cite{Vinod3} and study 
% viscosity and thermodynamic
% propeties ~\cite{Vinod4} of QGP in RHIC. Massive QPMs have been used to study thermodynamic behaviour of 
% QCD matter ~\cite{Salvatore}\cite{Gorenstein}\cite{Peshier} given by 
% LQCD data, interpreting SU(3) gauge field lattice data ~\cite{Politis} and shedding light on temperature 
% dependence of the thermal properties of matter and 
% transport coefficients ~\cite{Salvatore}\cite{Berrehrah}\cite{Vinod1} such as susceptibility, 
% shear viscosity and electrical conductivity.  

% Other attempts to construct quasi-particle models are by considering mass as a function of temperature[REF- 39-40(Peshier) Vinod Chandra's paper].
% The quasi-particle model[REF - Peshier] is introduced by interpreting quark-particle excitations as quasiparticles. This quasiparticle model is generalized to 
% non-zero quark chemical potential and in turn the quark masses depend on the chemical potential explicitly.  

After building the different quasi-particle models to 
map the QCD interaction, provided by LQCD data,
our next aim become to tune our models with experimental properties of QGP.
The experimental data~\cite{RHIC,LHC} from heavy ion collision experiments like
RHIC~\cite{RHIC} at BNL, USA and LHC~\cite{LHC} at CERN, Switzerland
indicate that QGP is a nearly perfect fluid system, which can be quantified by a very 
small values of shear viscosity to entropy density ratio $\eta/s$.
The entropy density $s$ of interacting QGP can already be known from LQCD direction,
from where the strength of thermodynamical phase-space part of shear viscosity $\eta$
can be fixed but it is relaxation time, which maps the collective dissipative properties
of QGP. We have explored this fact in present article. Mapping the interacting QCD picture
through different effective QCD models,
Refs~\cite{G_IFT,G_IFT2,G_CAPSS,Weise2,LKW,klevansky,klevansky2,Redlich_NPA,HMPQM1,HMPQM2,Deb,Kinkar_PNJL,Chowdhury,Marty} 
have addressed about the estimation of shear viscosity for quark matter. Whereas, different
quasi-particle models~\cite{Marty,Greco,Chandra_EPJC09_shear,Vinod-Suk1,Vinod-Suk2,Vinod-Suk3} 
provide the similar directional estimations. All those investigations have a common/main 
interest on searching some sources, for which QGP shows low viscous
behavior. Here we have particularly focus on the role of interaction for reduction of 
viscosity and fluidity of the medium, where interaction picture have been built via three different
simplified methodologies. We have also studied other transport coefficients like electrical conductivity 
of QGP.

The article is organized as follows. We have addressed the detail
methodologies for building the quasi particle models in next Sec.~(\ref{sec:Thermo}),
which are classified into three subsections - (\ref{sec:g}), (\ref{sec:Z}) and (\ref{sec:Gm})
to describe three different alternative ways to map interaction picture. After mapping the 
interaction in the models, we have applied to estimate transport coefficients of QGP, which 
are discussed in Sec.~(\ref{Tr_results}) and at the end, we have summarized our investigation.

% Hot QCD follows dissipative hydrodynamics during evolution in space-time. Knowledge of transport coefficients then are important to the understanding
% of the background physics ~\cite{Vinod-Suk1}. To incorporate the effects of a strongly interacting medium away from equilibriuma 
% quasi-particle description of hot QCD EOS has been employed through modelling of the momentum distribution of gluons and quarks  with non-trivial dispersio
% relations while extending the model for finite but small quark chemical potential. ~\cite{Vinod-Suk1} provides a way to visualize the effects of temperature
% on the transport coefficients due to effects of hot QCD. Quasi-particle description of hot QCD has also been utilized to study collective excitations
%  ~\cite{Vinod-Jamal}by modelling the effects through effective gluon and fugacities. 
% 
% Transport coefficients are also affected by by the size of the medium under consideration ~\cite{Kinkar}. At high temperature shear viscosity and
% thermal conductivity become independent of temperature while the bulk viscosity still depends on thelsize of the system. It has also been seen that transport
% coefficients are affected by the presence of a background field such as vector interaction ~\cite{Chowdhury}. This has been studied and compared
% in NJL and  EPNJL models. 

\section{Thermodynamic models by parameterizing LQCD data}
\label{sec:Thermo}
\subsection{Temperature dependent degeneracy factor}
\label{sec:g}
\begin{figure}
	\centering
	\includegraphics[scale=0.27]{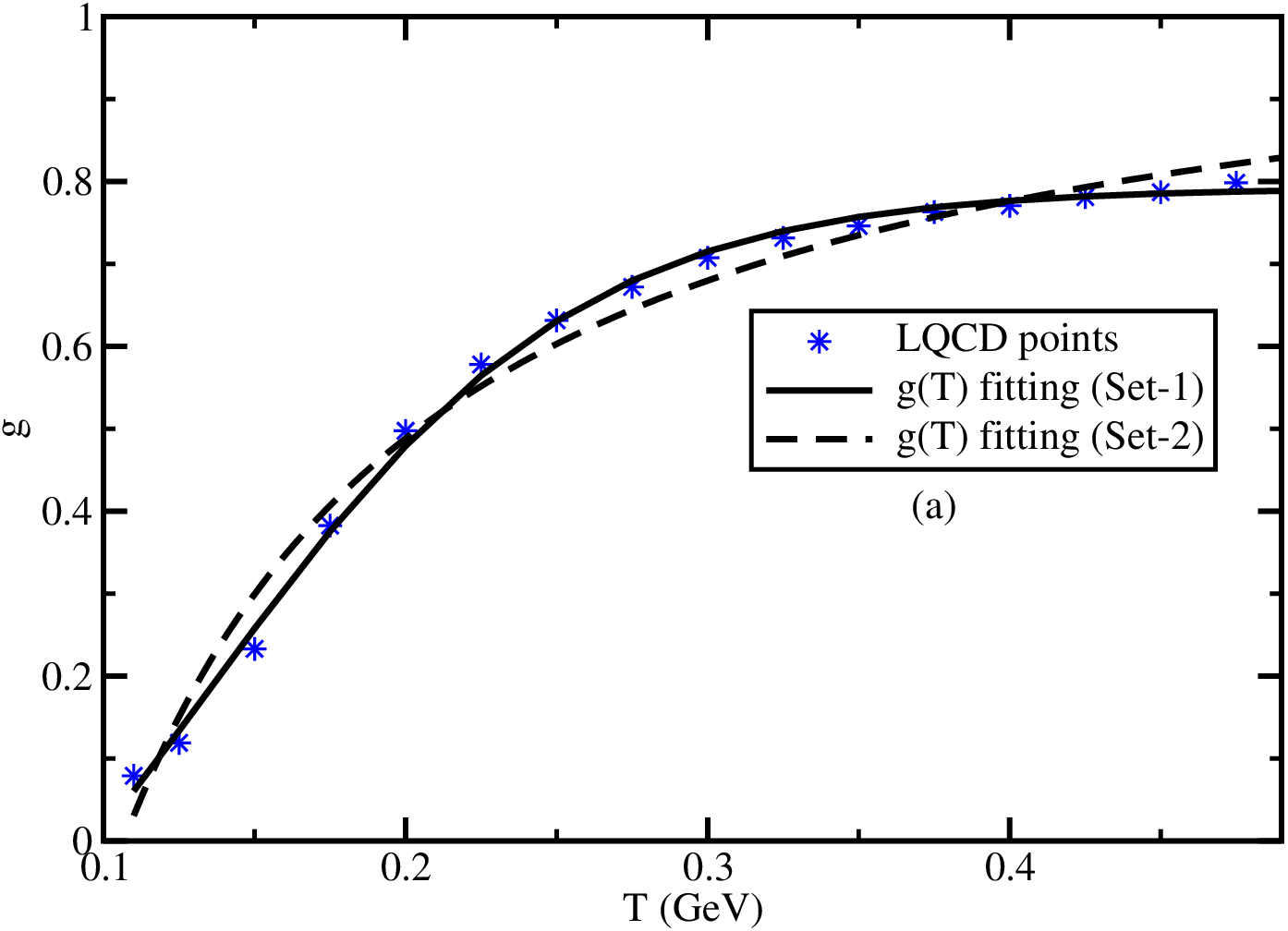} 
	\includegraphics[scale=0.27]{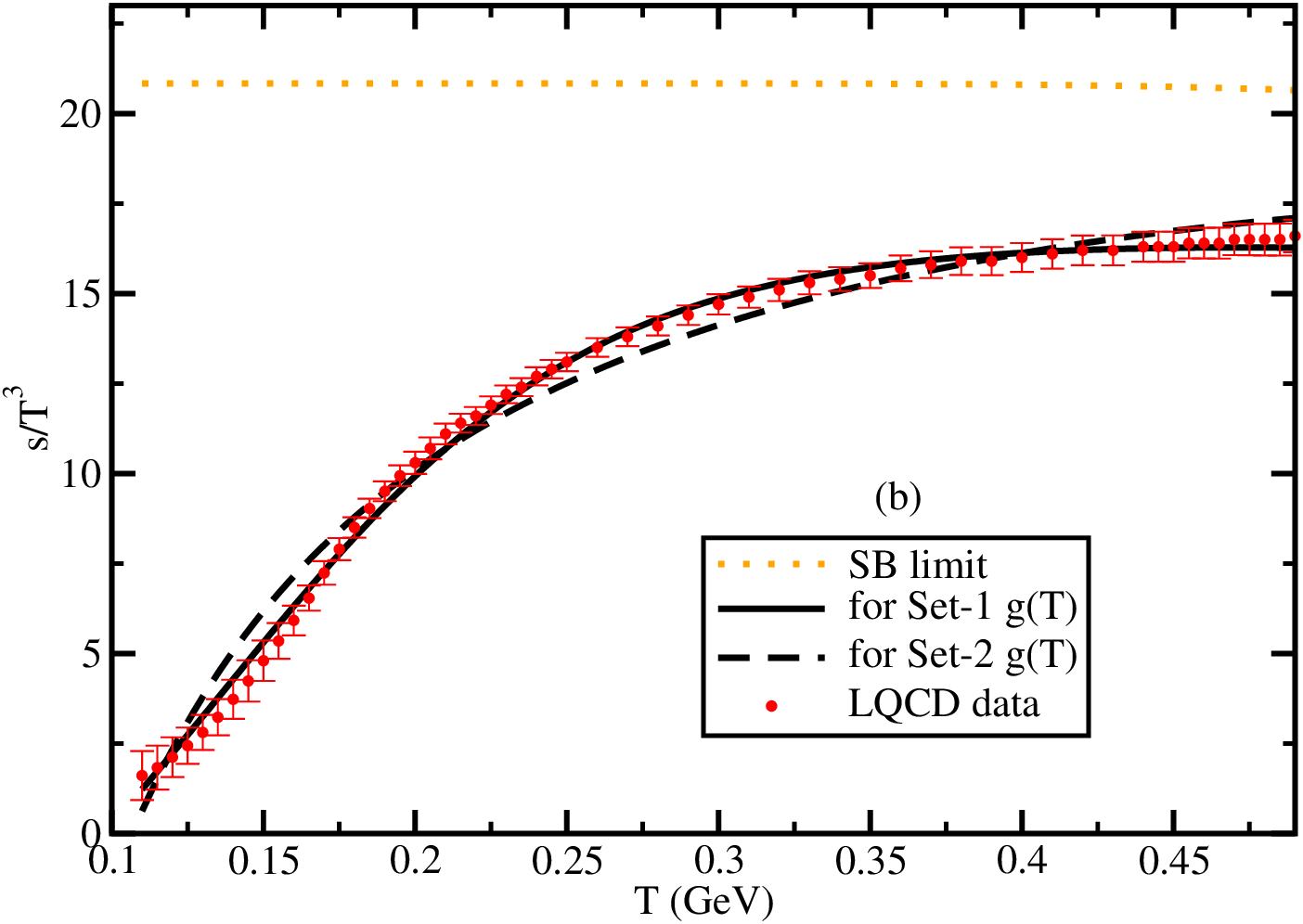} 
	\caption{(a) Temperature dependence degeneracy factors $g(T)$ parametrization curves - Set-1 (solid line),
	Set-2 (dash line) and LQCD extracted points (stars). (b) Their corresponding $s/T^3$ plots, where
	straight horizontal dotted line indicates SB limits of $s/T^3$.}
% 	\caption{(a) This graph illustrates the dependence of degeneracy of gluon and quark on temperature. 
% 	The property which the graph does not show is that, in high temperature-regime the 'g(T)' tends to unity as the 
% 	fitting curve of entropy density 's'should fit with the S-B limit. In this figure, the expected fit for degeneracy 
% 	curve can be seen with the limit as unity which goes with our theoretical background. By using this g(T), we can fit 
% 	the fitting curve of without any theoretical flaw and the fitting curve converges to the S-B limit. (b) This graph is 
% 	the fit of entropy density using the paramatrized expression g(T) i.e. degeneracy factor. 
% 	Here, the entropy density fit is decreasing slightly in the high temperature regime wrt the LQCD data. In the 2nd fit, 
% 	the fit approaching the S-B limit in the high temperature regime.    }
	\label{gvsT}
\end{figure}

Let us start with a non-interacting description of quark gluon plasma, where u, d, s quarks,
their anti-quarks and 8 different gluons are in thermal equilibrium. So quarks and anti-quarks
will follow Fermi-Dirac (FD) distribution and gluons obey Bose-Einstein (BE) distribution.
Following standard framework of statistical mechanics, one can calculate energy density 
and pressure of QGP system as
\be
\epsilon=g_g\int_{0}^{\infty}\frac{d^3p}{(2\pi)^3}\frac {p}{e^{\beta p}-1}
+g_u\int_{0}^{\infty}\frac{d^3p}{(2\pi)^3}\frac{\om_u}{e^{\beta\om_u}+1}
+g_s\int_{0}^{\infty}\frac{d^3p}{(2\pi)^3}\frac{\om_s}{e^{\beta\om_s}+1}
\label{e_QGP}
\ee
and
\bea
P &=& g_g\int_{0}^{\infty}\frac{d^3p}{(2\pi)^3}\Big(\frac{p}{3}\Big)\frac {1}{e^{\beta p}-1}
+g_u\int_{0}^{\infty}\frac{d^3p}{(2\pi)^3}\Big(\frac{p^2}{3\om_u}\Big)\frac{1}{e^{\beta\om_u}+1}
\nn\\
&+& g_s\int_{0}^{\infty}\frac{d^3p}{(2\pi)^3}\Big(\frac{p^2}{3\om_s}\Big)\frac{1}{e^{\beta\om_s}+1}~,
\label{P_QGP}
\eea
where $\om_u=\{p^2+m_u^2\}^{1/2}$, $\om_s=\{p^2+m_s^2\}^{1/2}$, with $m_u=0.005$ GeV, $m_s=0.100$ GeV.
The $g_u$, $g_s$ and $g_g$ are degeneracy factor u, d quarks, s quark and gluon respectively.
Their values are given below
\bea
g_u &=& ({\rm spin})\times({\rm particle/anti~particle})\times({\rm color})\times({\rm flavor})
=2\times 2\times 3\times 2=24~,
\nn\\
g_s &=& ({\rm spin})\times({\rm particle/anti~particle})\times({\rm color})\times({\rm flavor})
=2\times 2\times 3\times 1=12~,
\nn\\
g_g &=& ({\rm spin})\times({\rm flavor})
=2\times 8=16~.
\eea
Using thermodynamical relation with zero quark (and obviously gluon) chemical potential of QGP system, 
we can obtain entropy density
\be
 s=\frac{P+\epsilon}{T}~.
 \label{s_QGP}
\ee
Using Eqs.(\ref{e_QGP}), (\ref{P_QGP}), $s$ can be calculated and it will be very close to the 
analytic expression of Stephan-Boltzmann (SB) limit ($m_{u,s}\approx 0$):
\be
s=\Big[g_g+(g_u+g_s)\Big(\frac{7}{8}\Big)\Big]\frac{4\pi^2}{90}T^3\approx 20.8~T^3~.
\ee
According to lattice Quantum Chromo Dynamics (LQCD) calculation~\cite{LQCD1,LQCD2}, 
the numerical values of $s$ for QGP remain
always lower than its SB limits, which indicates about interaction picture of the system. 
For visualization, see Fig.~\ref{gvsT}(b), which will elaborately discussed latter.
Though LQCD is best tool to map the interaction
of QGP system, based on the theory of quantum chromo dynamics (QCD) but one can map this interaction via simple quasi-particle
model description. So, idea is to use simple and standard non-interacting thermodynamical relations, given in Eqs.~(\ref{e_QGP}), (\ref{P_QGP})
and (\ref{s_QGP}),  where a temperature dependent interacting information will carry by some quantity of quarks and gluons. 
One can get a huge number of references~\cite{NJL,NJL2,NJL3,Gorenstein,Peshier,Heinz,Bluhm}, 
where interaction information is mainly captured by temperature dependent of masses of quarks and gluons.
Here, we have caught the quantity degeneracy factors, whose temperature dependence 
can map the QCD interaction as an alternative and 
simplified way. From LQCD data~\cite{LQCD1,LQCD2}, we find that $s$ decreases if we decrease the temperature ($T$) and near 
the quark-hadron phase transition temperature ($T_c$),
its rate of decrement along -ve $T$-axis become maximum and at last in low temperature range, where QGP 
appears as hadronic degrees of freedom,
the values of $s$ becomes quite small. Now if one concentrate on hadronic temperature range ($T< T_c$), 
and try to get LQCD values
of $s(T< T_c)$, one of the immediate attempt comes through non-interacting hadronic matter (HM) calculation. 
Considering pion and Kaon
as most abundant mesons with u, d and s quarks, one can estimate $s(T< T_c)$ from the corresponding 
thermodynamical relations
\bea
s(T< T_c)&=&g_\pi\int_{0}^{\infty}\frac{d^3p}{(2\pi)^3}\Big[\om_\pi 
+\frac{\vp^2}{3\om_\pi}\Big]\frac{1}{e^{\beta\om_\pi}+1}
\nn\\
&+& g_K\int_{0}^{\infty}\frac{d^3p}{(2\pi)^3}\Big[\om_K +\frac{\vp^2}{3\om_K}\Big]\frac{1}{e^{\beta\om_K}+1}~,
\label{s_HM}
\eea
where $\om_\pi=\{\vp^2+m_\pi^2\}^{1/2}$, $\om_K=\{\vp^2+m_K^2\}^{1/2}$ with $m_\pi\approx 0.140$ GeV and $m_K=0.500$ GeV
and degeneracy factors of pion and kaon are $g_\pi=3$ and $g_K=4$. The numerical values of Eq.~(\ref{s_HM})
will be little below than its SB limits
\be
s(T< T_c)=(g_\pi +g_K)\frac{4\pi^2}{90}T^3\approx 3~T^3~.
\ee
The numbers of $s(T< T_c)$ in this simple non-interacting picture of HM system is quite close
to LQCD data around low temperature range of hadronic phase. So non-interacting QGP and HM system
provide us upper and lower estimations of $s$, within which LQCD data points~\cite{LQCD1,LQCD2} are located, where
major changes in values of $s$ (or any other thermodynamical quantities) are occurred near $T_c$. 
There are famous hadron reason gas (HRG) model~\cite{HRG_rev}, which can match LQCD data exactly
for $T\leq T_c$ range. Some Refs.~\cite{Anderson_HTL,Anderson_HTL2,Bannur} can cover
the $T\geq T_c$ range. While quasi-particle based 
Refs.~\cite{NJL,NJL2,NJL3,Chandra_PRC07,Chandra_EPJC09,Chandra_PRD11} 
attempted to span both temperature range $T\leq T_c$
and $T\geq T_c$ with QGP system, where $T\leq T_c$ zone basically tell about quasi particle nature 
of quarks and gluons inside hadrons. Here we also try to build that kind of quasi-particle identities
of quarks and gluons through entire temperature range. From simplified non-interacting picture,
we get a guidance that during the transition from quark to hadronic matter the degeneracy factors
has abruptly reduced from $(g_u, g_s, g_g)$ set to $(g_\pi, g_K)$ set, for which $s/T^3$ is reduced
from $20.8$ to $3$. Now LQCD data is saying that this reduction is not abruptly like first order phase transition 
rather smoothly like a cross-over transition. By assuming appropriate temperature dependent degeneracy
factors of quarks and gluons, one can construct LQCD data points for $s(T)$.
For this purpose, we have considered a temperature dependent factor $g(T)$, 
attached with $g_{q,s,g}$ and then match the LQCD 
data of $s(T)$~\cite{LQCD1,LQCD2}.
We get a parametrized expression:
\be
g(T)= a_0 - \frac{a_1}{e^{a_2(T-a_3)}+ a_4}~,
\label{gT_1}
\ee
where $a_0 = 0.793$, $a_1 = 0.687$, $a_2 = 16.284$, $a_3=0.170$, $a_4 = 0.560$~.
The above set of parameters (say set-1) provide better matching to LQCD data but it is not satisfying the expectation
of reaching SB limit of $s$ at $T\rightarrow\infty$.
To fulfill the condition, we have restricted $a_0=1$, and get the another parametrized function:
\be
g(T)= 1 - \frac{b_0}{e^{b_1(T-b_2)}+ b_3}~,
\label{gT_2}
\ee
where $b_0 = 0.793$, $b_1 = 0.687$, $b_2=0.170$, $b_3 = 16.284$, which can be called set-2.
% 
% g(T) maps the interaction between the QGP with the decreasing temperature. As, g(T) decreases with decrease in 
% temperature implies the degeneracy also decreases. 

Fig.~\ref{gvsT}(a) shows two set of $g(T)$ (dash and solid lines) and LQCD data points (stars)~\cite{LQCD1,LQCD2}. 
Their corresponding
values of $s/T^3$ is plotted in Fig.~\ref{gvsT}(b), where SB limit denoted by straight horizontal dotted line.
So effectively total degeneracy factor of QGP $g_u+g_s+g_g=52$ will be suppressed for considering $g(T)*g_{u,s,g}$,
and around $T=0.200$ GeV, $g(T)\approx 0.5$, which means effective degeneracy factor become $0.5\times 52=26$.
In hadronic temperature range, around $T\approx 0.120$ GeV, $g(T)\approx 0.13$ will provide effective degeneracy
factor $0.13\times 52=7$, which is exactly hadronic degeneracy factor $g_\pi +g_K=7$. So in this way, we might
roughly map QCD interaction picture via shrinking of degeneracy factor of quarks and gluons with lowering 
the temperature. This fact can be compared with the fact of temperature dependent 
degree of freedom for di-atomic or n-atomic molecule. At low temperature degrees of freedoms
of di-atomic or n-atomic molecules is $3\times 2-1$ or $3\times n - k$ because of its 1 or k
number of atomic bondings, which can be broken at high temperature and degrees of freedom enhanced
as
\bea
3\times 2-1=5 &\rightarrow& 3\times 2=6
\nn\\
{\rm or},
\nn\\
3\times n - k &\rightarrow& 3\times n~.
\eea
According to equipartition theorem of thermodynamics, internal energy of 
di-atomic or n-atomic molecular system will be proportional to its degrees 
of freedom, hence internal energy (other thermodynamical quantities) will also
be increased with increasing temperature.

\subsection{Temperature dependent Fugacity}
\label{sec:Z}
%
%
% In statistical mechanics, the macroscopic quantities can be calculated using the grand partition function,
% \begin{equation}
% \Theta=\sum_{0}^{\infty}z^{N}Q_N(V,T)
% \end{equation}
% where $z=e^{\beta\mu_{B}}$. $\mu_{B}$ is the chemical potential and $\beta=\frac{1}{k_{B}T}$. $Q_{N}(V,T)$ is the canonical partition function of N particles at temperature T and volume V, defined as
% \begin{equation}
% Q_N(V,T)=\sum_{i}e^{-\beta{E}}
% \end{equation}
% For Bose-Einstein system,
% \begin{equation}
% \Theta=\Pi_{i}\frac{1}{1-ze^{-\beta{\epsilon_{i}}}}
% \end{equation}
% For Fermi system,
% \begin{equation}
% \Theta=\Pi_{i}(1+ze^{-\beta{\epsilon_{i}}})
% \end{equation}
% According to Thermodynamics,
% \begin{equation}
% \Bigg(\frac{\delta{F}}{\delta{N}}\Bigg)_{V,T}=\mu_{B}
% \end{equation}
% Here $\mu_{B}$ is the chemical potential or the work done by the system in changing the number of states in the system. Since in our paper, we are working with a system where there is no change in the number of states, $\delta{N}=0$. Therefore $\mu_{B}=0$ and $z=1$. 
% \\
As an alternative method, instead of temperature dependent degeneracy factors of
quarks and gluons, one can mimic QCD interaction via temperature dependent fugacity $Z(T)$, 
as addressed in quasi-particle model of Chandra-Ravisankar~\cite{Chandra_PRC07,Chandra_EPJC09,Chandra_PRD11}.
This fugacity quantity is just for mimicking the QCD interaction but should not be confused
with fugacity $Z_{q,g}={\rm exp}(\mu_{q,g}/T)$ due to quark/gluon chemical potential, which
are considered as zero for present system.
% \\Recent experimental results indicate that the quark-gluon plasma (QGP) has already been produced at 
% RHIC and that its behavior is not close to that if an ideal gas. Measurements of flow parameters and 
% observations of jet quenching have stimulated the theoritical interpretation that the QGP behaves 
% like a nearly perfect fluid, characterized by a small value of viscosity to entropy density ratio, 
% lying in the range 0.1-0.3, the corresponding value of liquid helium(above superfluid transition temperature) 
% being 10. These observations signal the fact that the deconfined phase is strongly interacting and are 
% consistent with Lattice simulations, which predict a strongly interacting behaviour even at temperatures 
% much higher than $T_c$.
% To achieve these results, we have introduced Z in the distribution function. We have assumed z to be of the form
% \begin{equation}
% z=e^{\beta{\mu}}
% \end{equation}
% The fugacity of quark and gluon as a function of temperature has the same form.
% This is because the fitting values of the fugacity of gluon and quark obtained in the paper[C,R] are almost similar, 
% the equilibrium distribution function for quarks and gluons remain unchanged. 
% We have tried to map z to the interactions in the QGP-Hadron Matter system at zero chemical potential ($\mu_{B}=0$).
Hence, to get interacting values of $\ep$ and $P$, we have to replace 
$e^{\beta p}$,  $e^{\beta \om_u}$ and  $e^{\beta \om_s}$
in Eqs.~(\ref{e_QGP}), (\ref{P_QGP})
by $Z^{-1}e^{\beta p}$,  $Z^{-1}e^{\beta \om_u}$ and  $Z^{-1}e^{\beta \om_s}$, as
we are considering modified thermal distribution functions of u/d quark, s quark and gluon as
\bea
f_u &=& \frac{1}{Z^{-1}{\rm exp}\sqrt{p^2+m_u^2}\Big)+1}
\nn\\
f_s &=& \frac{1}{Z^{-1}{\rm exp}\Big(\beta \sqrt{p^2+m_s^2}\Big)+1}
\nn\\
f_g &=& \frac{1}{Z^{-1}{\rm exp}\Big(\beta p\Big)-1}~.
\eea
After knowing $\ep$, $P$, $s$ can be obtained from Eq.~(\ref{s_QGP}).
Keeping $Z$ as tuning parameter, we have matched the LQCD data~\cite{LQCD1,LQCD2} of $s$ and we get
a parametrized form
\be
 Z(T)=a_0-\frac{a_1}{e^{a_2(T-a_3)}+ a_4}~,
 \label{ZT_1}
\ee
Where $a_0 = 0.792535$, $a_1 = 0.686132$, $a_2 = 16.2834$, $a_3=0.170$ $a_4 = 0.56037$
At $T\rightarrow\infty$, $Z\rightarrow a_0$, which is 0.792535 but not 1. 
% At lower temperatures, the magnitude of both $Z_g$ and $Z_q$ is smaller indicating the 
% larger strength of interactions there[C,R]. In order to fit the LQCD data given to us, 
% we observe that even at extremely high temperatures, the aysmptotic limit does not tend to unity, 
It means that this set (say set-1) is not fulfilling the expectation of getting non-interacting
picture at high temperature limit.
% hence we are unable to explain the nature of interaction with the 
% obtained function of fugacity at high temperatures. 
% 
% If we use the fitting function of Z vs T as
%
\begin{figure}
\centering
\includegraphics[scale=0.27]{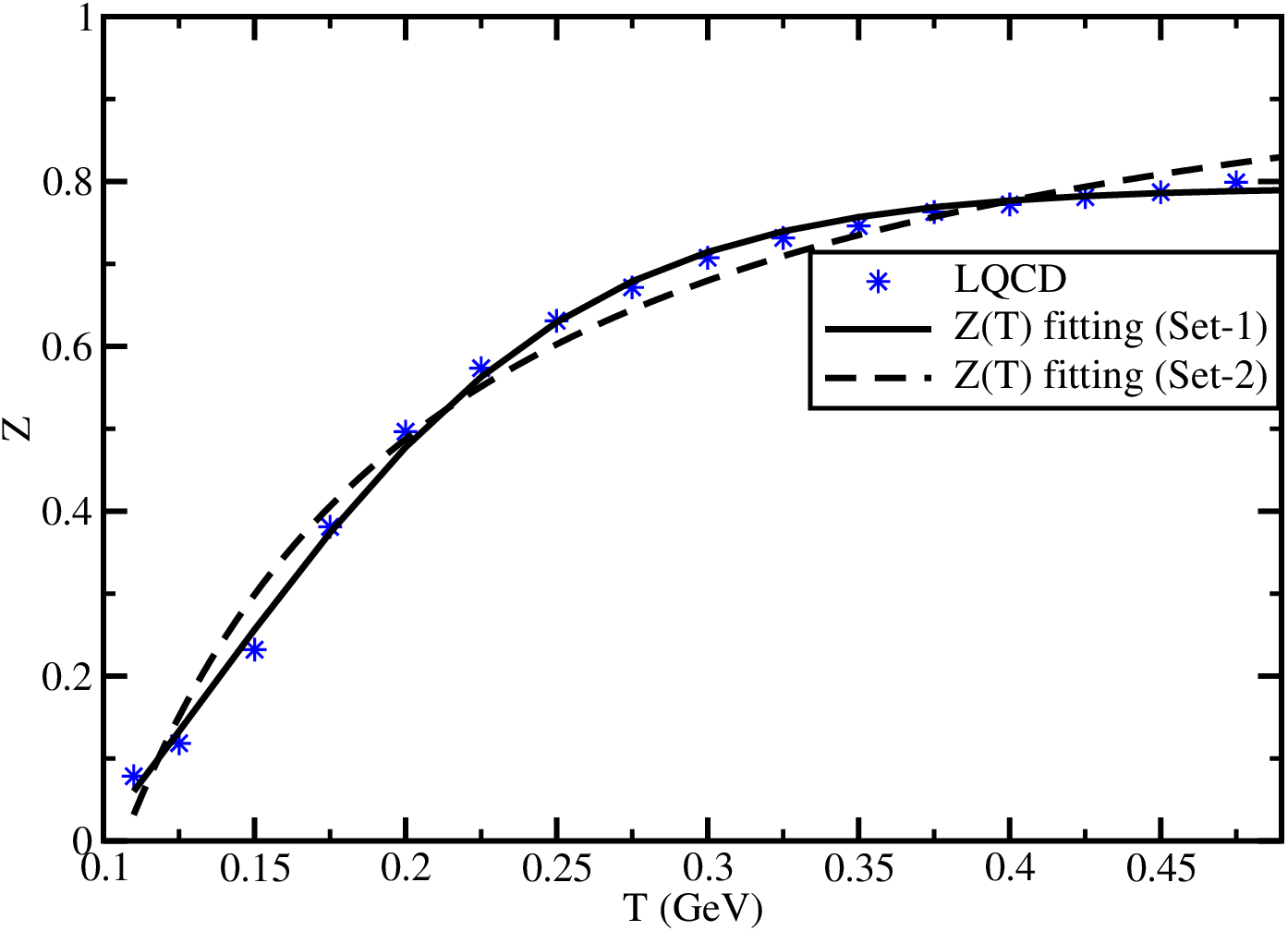} 
\includegraphics[scale=0.27]{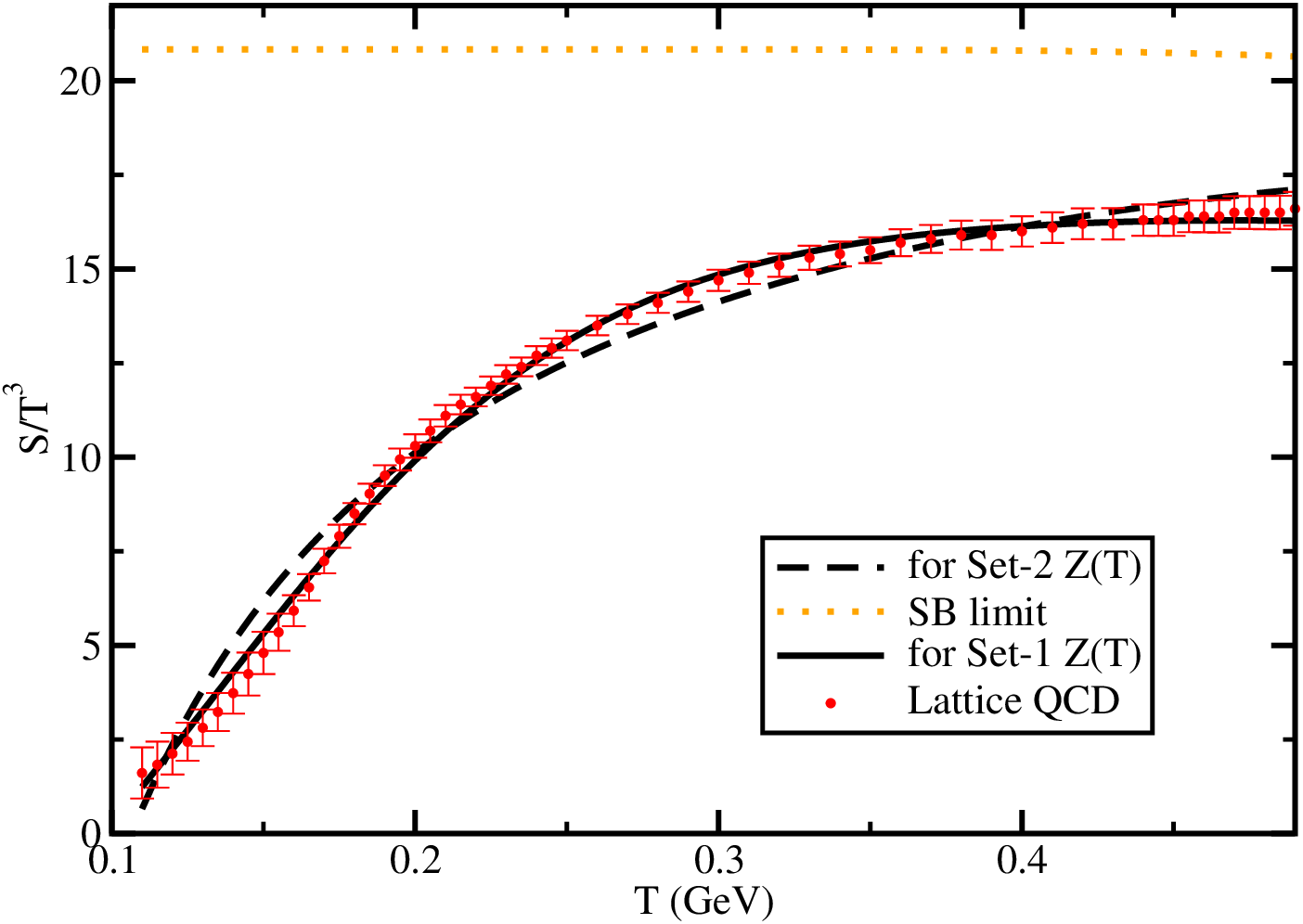} 
\caption{(a) Temperature dependence fugacity $Z(T)$ parametrization curves - Set-1 (solid line),
	Set-2 (dash line) and LQCD extracted points (stars). (b) Their corresponding $s/T^3$ plots, where
	straight horizontal dotted line indicates SB limits of $s/T^3$.}
\label{ZvsT}
\end{figure}
Therefore, restricting $a_0=1$, we find another parametrized form 
\be
Z=1-\frac{b_0}{e^{b_1(T-b_2)}+b_3}
\label{ZT_2}
\ee
Where $b_0=0.138935$, $b_1=1.445$, $b_2=0.170$ $b_3=-0.77362$ (say set-2)
and the expectation $Z(T\rightarrow \infty)\rightarrow 1$
% \be
% \lim_{T\rightarrow\infty}Z\rightarrow 1
% \ee
is well satisfied. These two sets of $Z(T)$ are shown 
by dash and solid curves in Fig.~\ref{ZvsT}(a), where stars
are LQCD data extracted points of $Z$. Their corresponding
values of $s/T^3$ is plotted in Fig.~\ref{ZvsT}(b), 
where SB limit denoted by straight horizontal dotted line.
%
% The fitting curve of entropy density does not follow the LQCD data exactly. 
% At low temperature, the magnitude of $Z$ is low, 
% implying that the interaction strength is high. 
% $Z$ increases with increase in temperature, indicating that interactions 
% in the system decrease with increasing temperature. We argue that as 
% temperature tends to infinity, fugacity reaches unity asymptotically. 
% At such high temperatures, no interactions prevail in the system and 
% all particles(quarks and gluons) exist as free particles.
%
\subsection{Temperature dependent Thermal width}
\label{sec:Gm}
Here, we will explore another alternative possibility to building a quasi-particle
model via temperature dependent thermal width of quarks and gluons. For this purpose,
let us revisit Eq.~(\ref{e_QGP}) and re-write in other way:
\bea
\epsilon &=& g_g\int_0^\infty dM\delta(M)
\int_{0}^{\infty}\frac{d^3p}{(2\pi)^3}\frac {\sqrt{p^2+M^2}}{{\rm exp}(\beta \sqrt{p^2+M^2})-1}
\nn\\
&+& g_u\int_0^\infty dM\delta(M-m_u)
\int_{0}^{\infty}\frac{d^3p}{(2\pi)^3}\frac{\sqrt{p^2+M^2}}{{\rm exp}(\beta \sqrt{p^2+M^2})+1}
\nn\\
&+& g_s\int_0^\infty dM\delta(M-m_s)
\int_{0}^{\infty}\frac{d^3p}{(2\pi)^3}\frac{\sqrt{p^2+M^2}}{{\rm exp}(\beta \sqrt{p^2+M^2})+1}~,
\label{e_QGP_Gm}
\eea
which is exactly same as Eq.~(\ref{e_QGP}), after using the identity $\int \delta(x-x_0)f(x)dx=f(x_0)$.
In non-interacting picture, quarks and gluons are stable and having delta function profile in mass space,
but for interacting case, delta function can be converted to Breit Weigner function 
\be
\rho(M)=\frac{1}{\pi}\Big(\frac{\Gamma_c}{\Gamma_c^2+(M-M_0)^2}\Big)~,
\ee
where $M$, $M_0$ is off-shell, on-shell mass of particle and $\Gamma_c$ is thermal width
of the particle, which basically maps the collision picture. One can get back delta distribution
for vanishing thermal width because of relation
\be
\delta(M-M_0)=\lim_{\Gamma_c\rightarrow 0}\rho(M)~.
\label{delta}
\ee
The transition from non-interacting to interacting picture via $\delta(M-M_0)\rightarrow\rho(M)$
in energy/momentum space will be more clear through the transition $1\rightarrow e^{-t/\tau_c}$ or
$1\rightarrow e^{-x/\lambda_c}$ in time/position space, where $\tau_c=1/\Gamma_c$ or $\lambda_c\propto 1/\Gamma_c$ 
is mean collisional time or mean free path of the interacting medium. 
It is Fourier's transformation, which can connect between $M$-axis and $t$-axis, 
hence, $\delta(M-M_0)$ and $1$ are linked in non-interacting picture, while
$\rho(M)$ and $e^{-t/\tau_c}$ are linked in interacting picture.
We see that
due to interaction, finite $\Gamma_c$, $\tau_c$, $\lambda_c$ are expected and therefore,
probability of particle with time/position will be exponentially reduced. Similar to
limiting case (\ref{delta}), one can get back constant probability in time/position
space for $\Gamma_c\rightarrow 0$ as
\be
\lim_{\Gamma_c\rightarrow 0}e^{-t/\tau_c}\rightarrow 1~.
\label{GmT}
\ee
So, transforming the expressions from $\Gamma_c\rightarrow 0$ or $\tau_c\rightarrow\infty$
to finite $\Gamma_c$ or $\tau_c$, one can build non-interacting to interacting picture description.
Following that, if delta functions in Eq.~(\ref{e_QGP_Gm}) will be replaced by their corresponding $\rho(M)$'s
with $M_0=0,m_u,m_s$ for gluon, u/d quark and s quark, then non-interacting to interacting energy density
expressions can be obtained. Applying same technique to Eq.~(\ref{P_QGP}) and then in Eq.~(\ref{s_QGP}),
one finally get $s$ of interacting system, where $\Gamma_c$ parameter can be tuned to match LQCD data~\cite{LQCD1,LQCD2}.
\begin{figure}
\centering
\includegraphics[scale=0.27]{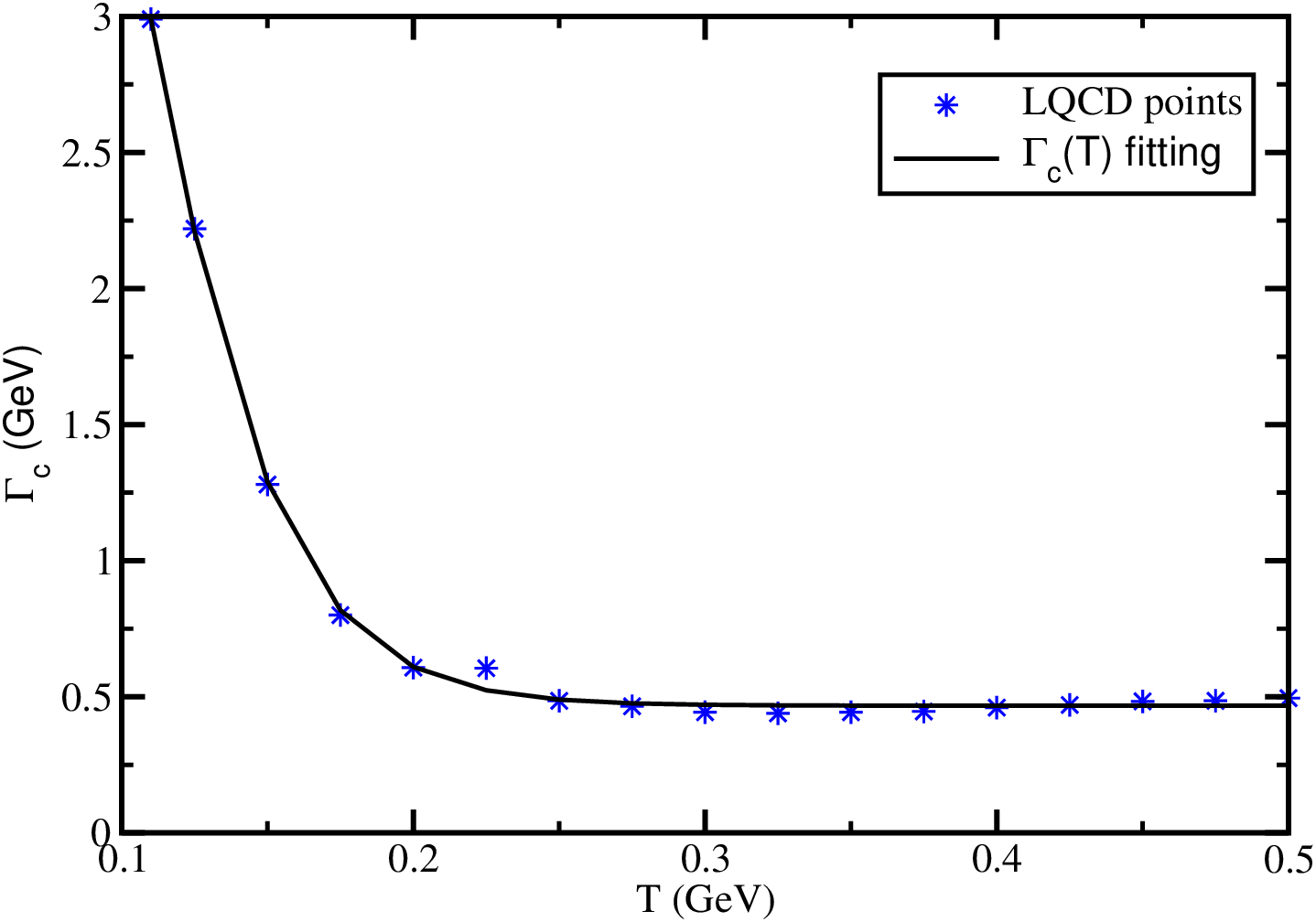}
\includegraphics[scale=0.27]{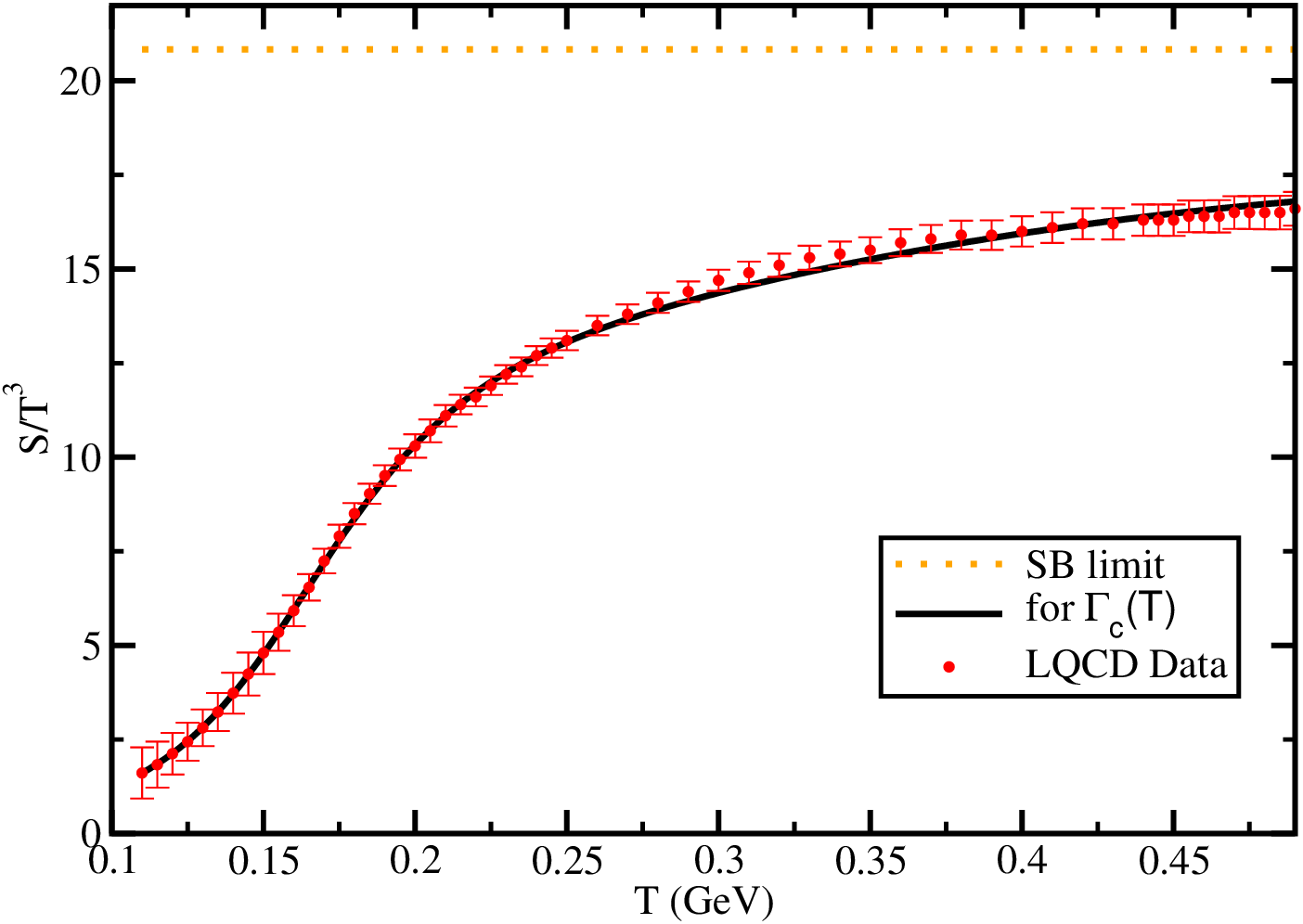} 
\caption{(a) Temperature dependence thermal width $\Gamma_c(T)$ parametrization curve (solid line)
	and LQCD extracted points (stars). (b) Their corresponding $s/T^3$ plots, where
	straight horizontal dotted line indicates SB limits of $s/T^3$.}
\label{GammavsT}
\end{figure}
By this matching, we get parametrized of $\Gamma_c(T)$:
\be
\Gamma_c(T)=a_0-\frac{a_1}{e^{a_2(T-a_3)}+ a_4}
\label{Gm_fit}
\ee
where, $a_0=6.76802$, $a_1=88.6265$, $a_2=-37.3715$, $a_3=0.170$, $a_4=14.0653$.
It is plotted in Fig.~\ref{GammavsT}(a) and corresponding $s/T^3$
is displayed in Fig.~\ref{GammavsT}(b).
In hadronic temperature range, $\Gamma_c(T)$ decreases with $T$ but it
saturates with the values $\Gamma_c\approx 0.500$ GeV in quark temperature domain.
Here, one can again find set-2 parametrization of $\Gamma_c(T)$, where $\Gamma_c\rightarrow 0$
at $T\rightarrow\infty$ but that choice provide a very bad matching of LQCD data, so we
have not considered that set. 
% 
% comes to be a decreasing function with increasing 
% temperature when it is used to fit the LQCD data, converging to a value 
% at high temperature regime representing QGP plasma state. As, the temperature 
% starts increasing $\Gamma(T)$ starts increasing representing increasing interations 
% in QGP state forming new Hadronic matter. $2\Gamma$ represents the width of the 
% distribution which implies with decreasing temperature the width of the distribution 
% gets wider which represents many masses have their contribution in the Hadronic matter. 
% In the high temperature regime, $\Gamma$ is small, representing contribution of only 
% quarks and gluons mass in th QGP plasma state.    

%
\section{Estimation of shear viscosity and electrical conductivity}
\label{Tr_results}
\begin{figure}
	\centering
	\includegraphics[scale=0.27]{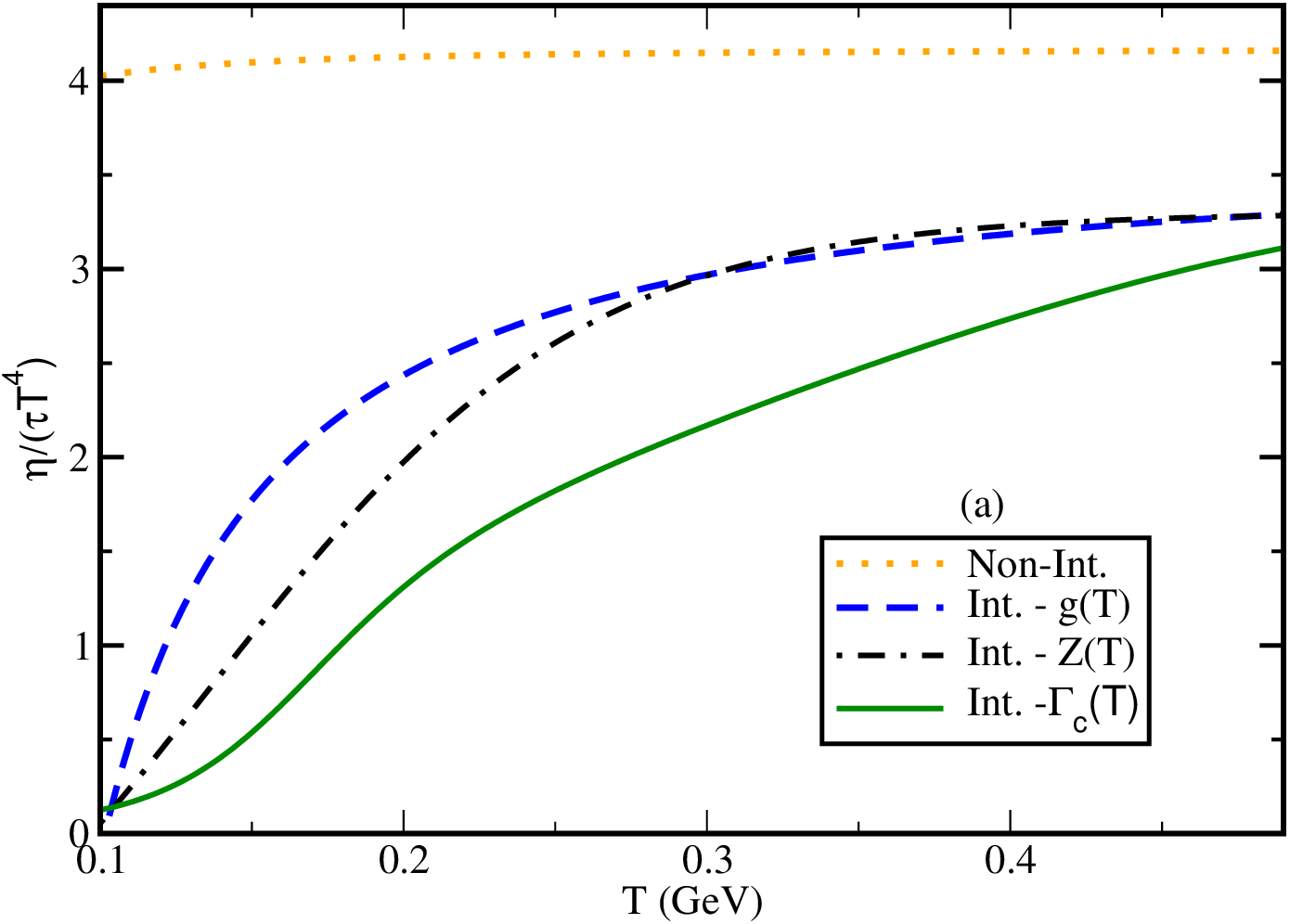} 
	\includegraphics[scale=0.27]{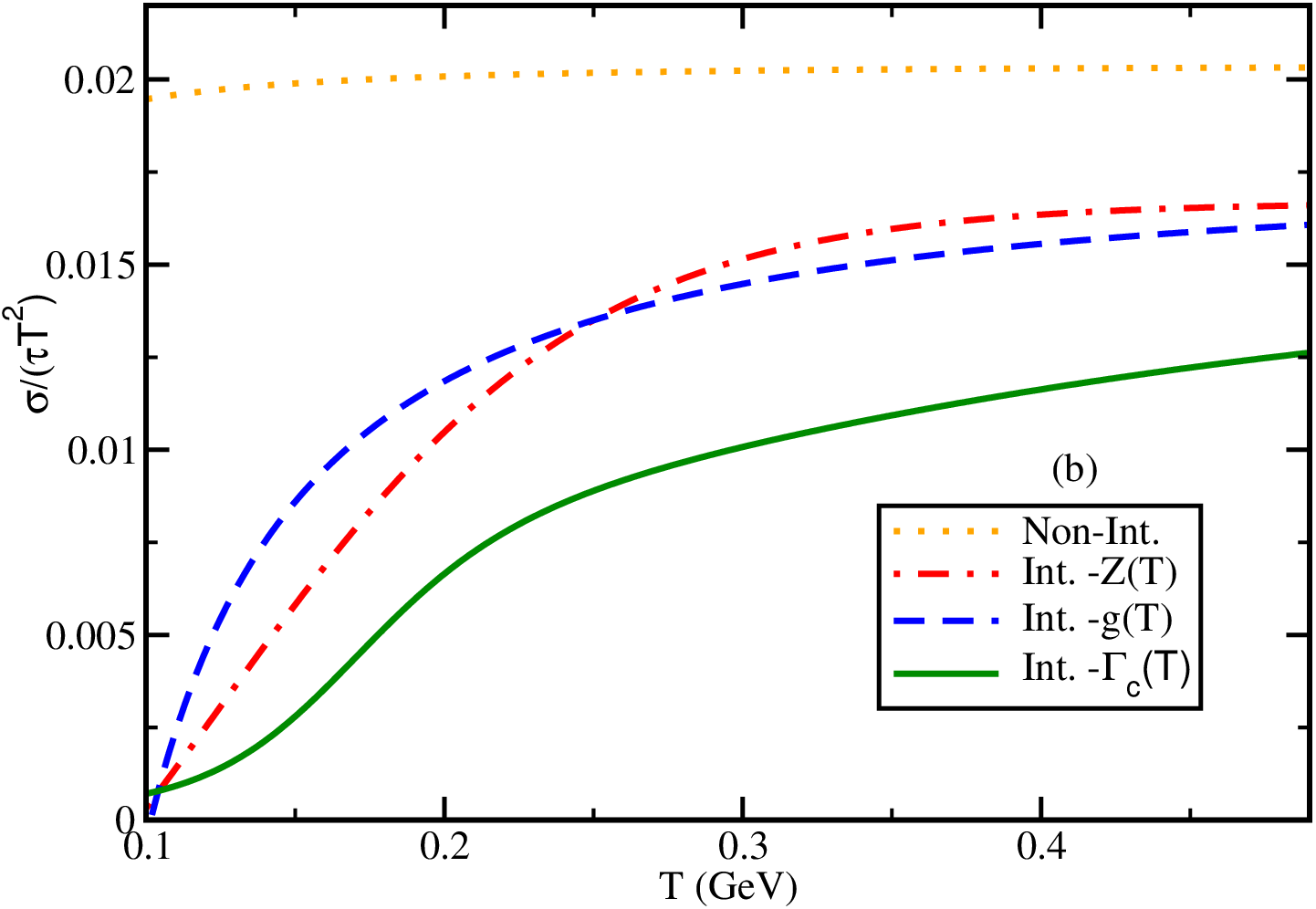} 
	\caption{Temperature dependent of (a) $\eta/(\tau T^4)$, (b) $\sigma/(\tau T^2)$ are plotted by using temperature dependent 
	degeneracy factor $g(T)$ (dash line), fugacity $Z(T)$ (dash-dotted line) and thermal width $\Gamma_c(T)$ (solid line). Horizontal
	dotted lines indicate corresponding non-interacting values.}
	\label{EtaSig}
\end{figure}

After developing different alternative ways to build a quasi-particle model, which
match well the LQCD data of QGP thermodynamics, here, we will plug in that 
QCD interaction in calculations of different transport coefficients like shear viscosity
($\eta$) and electrical conductivity ($\sigma$). We know that quasi-particle expressions
of $\eta$ and $\sigma$, addressed in Appendix/Sec.~(\ref{App}), can be obtained from
either relaxation time approximation (RTA) of kinetic theory~\cite{Gavin,chakrabortty,Chowdhury} or 
from one-loop diagram of respective correlators, based on Kubo relations~\cite{Nicola,Jeon,G_Kubo,G_el}. 
Interestingly both methodology provide
same final expressions for different transport coefficients because of their own approximations.
%Kinetic theory is approximated by crude relaxation time scale in relativistic Boltzmann
%equation, although one can go through its (possible) general aspects. 
The detail derivation of the final expressions of $\eta$, $\sigma$ from two methodology are
described in Appendix/Sec.~(\ref{App}). Let us start here from directly final expressions for 
QGP system,
\bea
\eta_{\rm t}&=&\sum_{i=g,u,s}\eta_{i}
\nn\\
&=& \frac{g_{g}}{15T} \int\frac{d^3\vp}{(2\pi)^3} 
\vp^2 \tau\: f(1 + f) 
+\frac{g_u}{15T} \int\frac{d^3\vp}{(2\pi)^3} \left(\frac{\vp^2}{\om_u}\right)^2 \tau\: f(1- f) 
\nn\\
&&~~~~~~~~+\frac{g_s}{15T} \int\frac{d^3\vp}{(2\pi)^3} \left(\frac{\vp^2}{\om_s}\right)^2 \tau\: f(1- f)
\label{eta_QGP}
\\
\sigma_{\rm t} &=& \sum_{i=u,s}\sigma_{i}
\nn\\
&=& \frac{e^2_u g_e}{3T} \int{\frac{d^3\vp}{(2\pi)^3} \left(\frac{\vp}{\om_u}\right)^2} \tau \: f(1 - f) 
+\frac{e^2_s g_e}{3T} \int{\frac{d^3\vp}{(2\pi)^3} \left(\frac{\vp}{\om_s}\right)^2} \tau \: f(1 - f)~.
\label{el_QGP}
\eea
Gluons are uncharged and hence do not contribute to electrical conductivity.
The above the Eqs.~(\ref{eta_QGP}), (\ref{el_QGP}) in massless limit (i.e. $m_{u,s}\rightarrow 0$)
will covert to a simple analytic function,
\bea
\eta_{\rm t} &=& \Big[g_g + \Big(\frac{7}{8}\Big)(g_u+g_s)\Big]\frac{4\tau}{5\pi^2}\zeta(4)T^4
= \Big[16 + \Big(\frac{7}{8}\Big)(24+12)\Big]\frac{4\tau}{5\pi^2}\zeta(4)T^4
\nn\\
\sigma_{\rm t} &=& (e_u^2g_{e}+e^2_sg_{e})\frac{\tau}{3\pi^2}\zeta(2)T^2
= \Big[12\Big(\frac{5e^2}{9}\Big)+12\Big(\frac{e^2}{9}\Big)\Big]\frac{\tau}{3\pi^2}\zeta(2)T^2
\eea
where $\zeta(4)=\pi^4/90$, $\zeta(2)=\pi^2/6$. So we will get massless limit
of transport coefficients, $\eta_{\rm t}/(\tau T^4)\approx 4.16$
and $\sigma_{\rm t}/(2\tau T^2)\approx 0.22e^2\approx0.02$ as we have found 
similar type (SB) limiting values for thermodynamical quantity $s/T^3\approx 20.8$.
We have plotted $\eta_{\rm t}/(\tau T^4)$ vs $T$ and $\sigma_{\rm t}/(2\tau T^2)$ vs $T$ 
curves (brown dotted line) in Fig.~\ref{EtaSig}(a) and (b), which should be exactly
straight horizontal line for massless case but actually it will not because of finite
values of $m_u$ and $m_s$. So, for non-interacting picture, we are getting roughly
constant values of $\eta_{\rm t}/(\tau T^4)$ and $\sigma_{\rm t}/(2\tau T^2)$, which 
can be modified in interacting picture. As we noticed in earlier section,
we have attempted to build interaction picture of QGP system by introducing
temperature dependent (1) degeneracy factor $g(T)$, (2) fugacity $Z(T)$ and
(3) thermal width $\Gamma_c(T)$. Implementing them one by one, we will estimate their
corresponding values of transport coefficients.

Let us first come to temperature dependent degeneracy factor case.
Multiplying $g(T)$ of Eq.~(\ref{gT_1}) with $g_{g,u,s}$ in 
Eqs.(\ref{eta_QGP}) and (\ref{el_QGP})~, we have obtained the curves, shown
by dash line in Fig.~\ref{EtaSig}(a) and (b). Next, we go to the $Z(T)$ case,
where we have modified the distribution function by using $Z(T)$ and got
dash-dotted curves in Fig.~\ref{EtaSig}(a) and (b). Then we have done the third case
$\Gamma_c(T)$. Here, we have to first replace energy of gluons, quarks as $\sqrt{\vp^2+M^2}$
in Eqs.~(\ref{eta_QGP}) and (\ref{el_QGP}) and then we have to integrate by their respective
spectral function $\rho(M)$'s. So the modified expressions of Eqs.~(\ref{eta_QGP}) and (\ref{el_QGP})
will be
\bea
\eta_{\rm t}&=&\sum_{i=g,u,s}\eta_{i}
\nn\\
&=& \frac{g_{g}}{15T}\int\frac{dM}{\pi}\Big(\frac{\Gamma_c}{\Gamma_c^2+M^2}\Big) \int\frac{d^3\vp}{(2\pi)^3} 
\frac{\vp^4}{\vp^2+M^2} \tau\: f(1 + f) 
\nn\\
&+&\frac{g_u}{15T} \int\frac{dM}{\pi}\Big(\frac{\Gamma_c}{\Gamma_c^2+(M-m_u)^2}\Big)\int\frac{d^3\vp}{(2\pi)^3} \frac{\vp^4}{\vp^2+M^2} \tau\: f(1- f) 
\nn\\
&+&\frac{g_s}{15T} \int\frac{dM}{\pi}\Big(\frac{\Gamma_c}{\Gamma_c^2+(M-m_s)^2}\Big)\int\frac{d^3\vp}{(2\pi)^3} \frac{\vp^4}{\vp^2+M^2} \tau\: f(1- f)
\\
\label{eta_QGP_gm}
\sigma_{\rm t} &=& \sum_{i=u,s}\sigma_{i}
\nn\\
&=& \frac{e^2_u g_e}{3T} \int\frac{dM}{\pi}\Big(\frac{\Gamma_c}{\Gamma_c^2+(M-m_u)^2}\Big)\int\frac{d^3\vp}{(2\pi)^3} \frac{\vp^2}{\vp^2+M^2} \tau \: f(1 - f) 
\nn\\
&+& \frac{e^2_s g_e}{3T} \int\frac{dM}{\pi}\Big(\frac{\Gamma_c}{\Gamma_c^2+(M-m_s)^2}\Big)
\int\frac{d^3\vp}{(2\pi)^3} \frac{\vp^2}{\vp^2+M^2} \tau \: f(1 - f)~.
\label{el_QGP_gm}
\eea
The green solid lines in Figs.~\ref{EtaSig}(a) and (b) is showing the respective transport coefficients curves, where
QCD interaction has been mapped through temperature dependent thermal width, given in Eq.~(\ref{Gm_fit}). So from the 
dash, dash-dotted and solid line curves of Figs.~\ref{EtaSig}(a) and (b), based on $g(T)$, $Z(T)$ and $\Gamma_c(T)$
parameterizing quasi-particle models, we notice a common qualitative message - transport coefficients of non-interacting
QGP will be reduced because of interaction as noticed in thermodynamical quantity like $s$.
\begin{figure}
	\centering
	\includegraphics[scale=0.27]{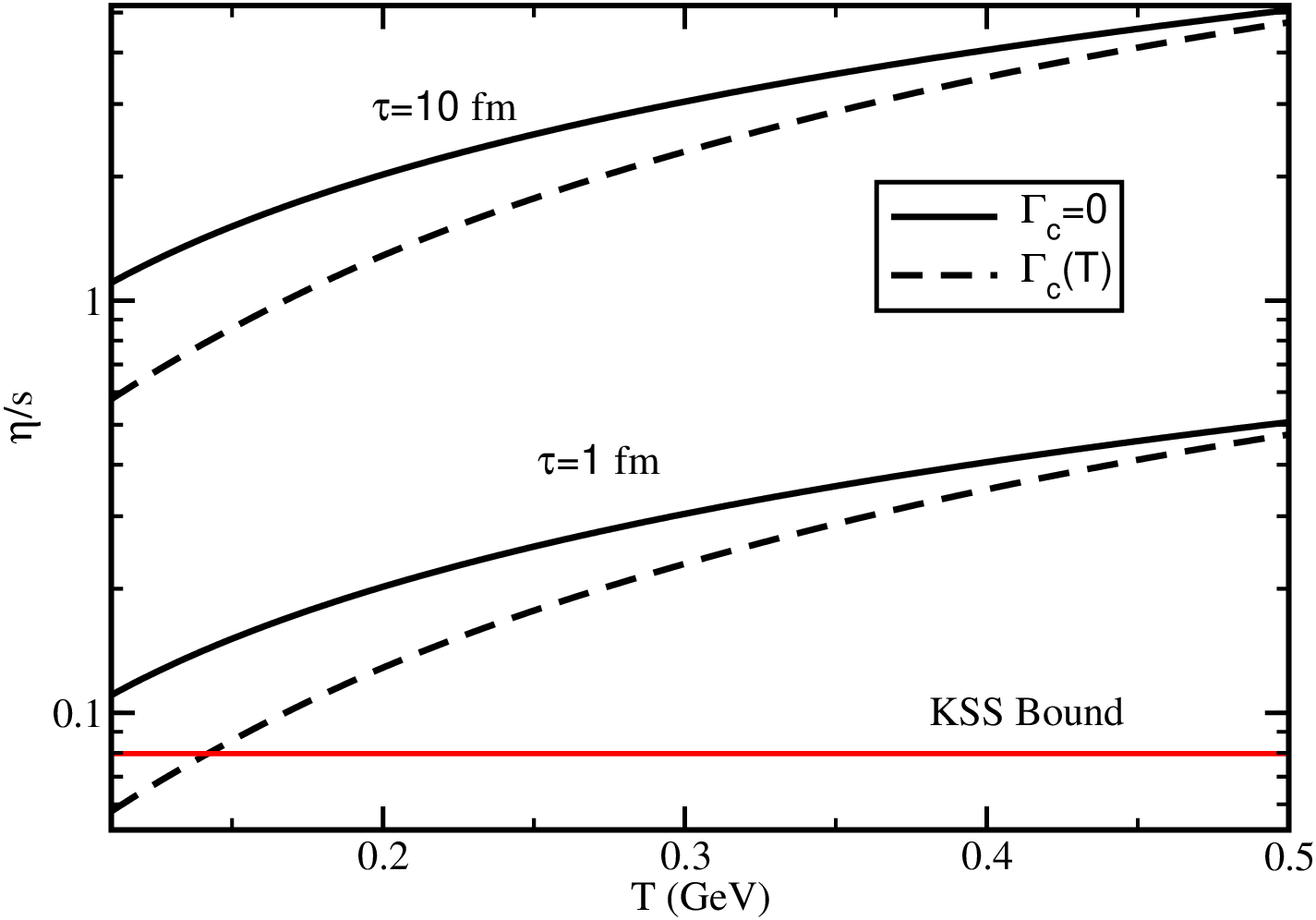} 
	\includegraphics[scale=0.27]{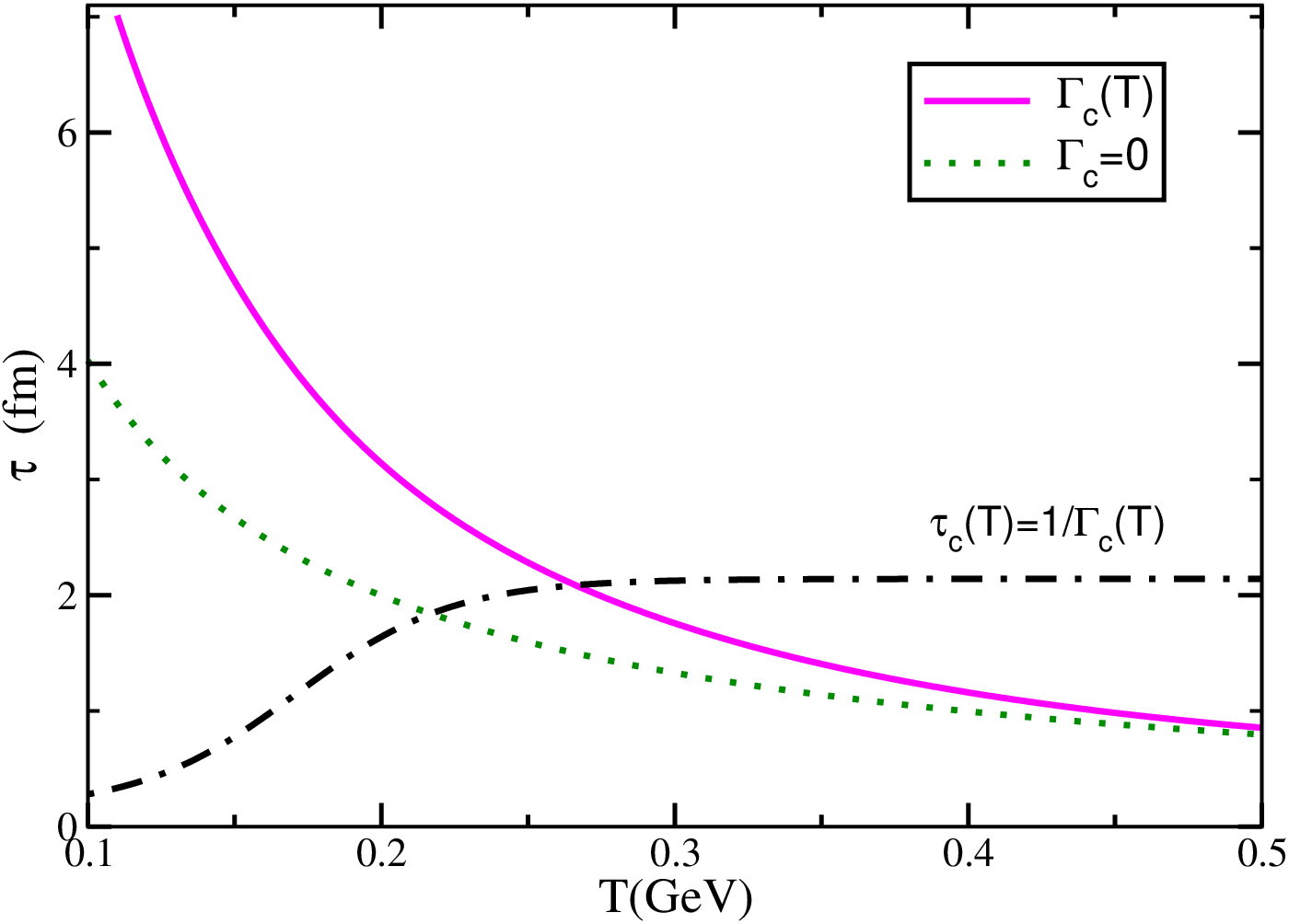} 
	\caption{(a) $\eta/s$ vs $T$ for non-interacting (solid line) and interacting (dash line) cases by considering
	$\Gamma_c=0$ and $\Gamma_c(T)$. Curves are plotted for plotted for two values of relaxation time. (b) By imposing
	$\eta/s=1/(4\pi)$, $\tau(T)$ has been found for non-interacting or $\Gamma_c=0$ (dotted line) and interacting or
        $\Gamma_c(T)$ (solid line) cases. The $\tau_c(T)=1/\Gamma_c(T)$ is also plotted to compare with relaxation time scale $\tau$.}
	\label{EtaStauC}
\end{figure}
%

% as obtained from numerical simulations from thermodynamics 
% to fit the LQCD data, to $\eta$, and plotting $\frac{\eta}{\tau_{\pi}T^4}$ indicates that the 
% viscosity of interacting systems lies well below non-interacting systems, i.e. g(T)=1 and Z(T)=1. 
% This also follows from the experimental data. The Quark-Gluon Plasma system consists of up, down, 
% strange quarks and gluons and all the species contribute to the viscosity of the medium.
% 
% Putting the same g(T), Z(T) and $\Gamma$(T) in $\sigma$, and plotting $\frac{\sigma}{\tau_{\pi}T^2}$, 
% we observe that the electrical conductivity of interacting systems is lower than the electrical 
% conductivity of non-interacting systems, as shown by experimental data.
% The Quark-Gluon plasms system consists of up, down, strange quarks and gluons, but only quarks are 
% charged and hence the quark medium has electrical conductivity. 

Seeing the reduction in thermodynamical phase-space of shear viscosity in interaction picture, we
have inclined to search - is there any dominant/partial/non-negligible role of interaction for reducing fluidity
of QGP, for which we are getting a (nearly) perfect fluid system? However, instead of $\eta$ only, we should found 
the dimensionless ratio $\eta/s$, which basically measures the fluidity of a system. Since $\eta$ and $s$ both are
facing reduction due to interaction, so it is not straight forward to make a comment about the role of interaction
on QGP fluidity.  
% 
% Kovtun et al showed that certain special field theories [anti-deSitter/conformal 
% field theory (AdS/CFT)] that are dual to black branes in higher space-time dimensions 
% have the ratio of shear viscosity to entropy density $\frac{\eta}{s}=\frac{1}{4\pi}$ 
% (in natural units with $\hbar = k_B = c = 1$). This limit is called the KSS Bound. 
% They argued that any substance should have a lower bound on $\frac{\eta}{s}$ because of 
% the uncertainty principle. Substance which have $\frac{\eta}{s}=\frac{1}{4\pi}$ are 
% classified as perfect fluids. On making precise calculations of the ratio of $\frac{\eta}{s}$ as 
% functions of temperature, we notice that for collisional time, $\tau_{\pi}=1 fm$, $\frac{\eta}{s}$ 
% for an interacting medium crosses the lower limit of $\frac{1}{4\pi}$ for $\Gamma$(T). 
% At temperatures below 0.15 GeV, where the $\frac{\eta}{s}$ for $\Gamma$(T) goes below the 
% KSS Bound, we argue that the theory we used to find $\Gamma$(T) is invalid. 
We have plotted first $\eta/s$ for non-interaction system by using $\tau=1$ and $10$ fm, shown by
solid line in Fig.~\ref{EtaStauC}(a). Then generated same curves for $g(T)$, $Z(T)$ based quasi-particle
model and interestingly, we have found those interacting curves are exactly coincided with non-interacting curves.
It reflects that the quantity $g(T)$ and $Z(T)$ in $\eta$, $s$, which map QCD interaction, are exactly
canceled in ratio $\eta/s$. Therefore, $\eta/s$ for non-interacting and interacting QGP system,
based on $g(T)$ and $Z(T)$ parametrization are exactly same. However, this picture is not true in $\Gamma_c(T)$
parametrization. The dash lines in Fig.~\ref{EtaStauC}(a) show the interacting curves of $\eta/s$, based on $\Gamma_c(T)$
parametrization for $\tau=1$ and $10$ fm. We notice a clear reduction of $\eta/s$ i.e. 
fluidity of QGP because of interaction.

Now, relaxation time $\tau$ in the expression of shear viscosity and electrical conductivity
keep as free parameter and in principle it is different from particle collisional time $\tau_c=1/\Gamma_c$.
By definition, relaxation time is the time scale, required to return from non-equilibrium to equilibrium
distribution function. Now, this non-equilibrium state can be created either by shear stress or by electrical
field and because of different sources of dissipative forces, relaxation time of different transport
coefficients might be different. However, for simplicity we consider that all of them are same. 
Now this relaxation time $\tau$ for shear stress can map the interaction of dissipative QGP fluid while
quasi-particle collisional time $\tau_c$ is mapping LQCD thermodynamic. To search connection between them,
we have first plotted $\tau_c(T)$ by dash-dotted line in Fig.~\ref{EtaStauC}(b). While, free parameter
$\tau$ in $\eta$ can be guessed from experimental data of QGP fluid, which indicates about its perfect fluid
nature i.e. $\eta/s$ touch the KSS value $1/(4\pi)$. So
imposing $\eta/s=1/(4\pi)$ for non-interacting and interacting picture, based on $\Gamma_c(T)$ parametrization,
we have generated dotted and solid lines respectively in Fig.~\ref{EtaStauC}(b). For non-interacting
and massless case (i.e. $m_{u,s}\rightarrow 0$), we get simplest relation 
\bea
\frac{\eta}{s}&=&\frac{1}{4\pi}
\nn\\
\Rightarrow \frac{\tau T}{5}&=&\frac{1}{4\pi}
\nn\\
\Rightarrow \tau &=&\frac{5}{4\pi T}~,
\eea
although it will be little different for $m_u=0.005$ GeV and $m_s=0.100$ GeV. For interaction
case, we will not get any analytic expression as a clue, rather we have to follow the numerical
values of solid lines of Fig.~\ref{EtaStauC}(b) to understand the trend.

Let us try to relate these two time scale
roughly as $\tau=\phi(T)\tau_c$, where $\phi(T)< 1$ for quark temperature domain and
$\phi(T)> 1$ for hadronic temperature domain are noticed in Fig.~\ref{EtaStauC}(b). 
If we roughly understand larger time scale
as more macroscopic, then at quark temperature domain, $\tau_c$ is appeared as macroscopic
scale, whereas at hadronic temperature domain $\tau$ plays the macroscopic role. In general,
for good kinetic theory approximation, we consider $\tau\approx\tau_c$, which might be
more or less applicable near and above transition temperature at least in order of 
magnitude ($\tau_c\approx2$ fm, $\tau\approx 1$ fm). It indicates that high temperature 
QCD interaction time scale, covered by LQCD data is quite well agreement with shear dissipative 
interaction of QGP, although this equivalence might not be true in low temperature hadronic phase. 
%
% For non-interacting species, i.e. $g(T)=1,Z(T)=1$ and for interacting species, i.e. g(T) 
% and Z(T), the plot of $\frac{\eta}{s}$ lies above the KSS bound for collisional time $\tau_{\pi}=1 fm$.
%
% For both interacting species, g(T), Z(T) and $\Gamma$(T) and non-interacting 
% species, g(T) = Z(T) = 1, the plot of $\frac{\eta}{s}$ lies above the KSS bound for 
% collisional time $\tau_{\pi}=10 fm$.
% \\
% We notice that with linear increase in collisional time $\tau_{\pi}$, 
% the plot of $\frac{\eta}{s}$ exhibits a linear increase. This is due to 
% \begin{equation}
%     \frac{\eta}{s}=\frac{g\int_{0}^{\infty}\frac{d^3p}{2\pi^3}
%     \frac{\vp^{4}}{E^2}\tau_{\pi}f_{o}(1-f_{o})}{g\int_{0}^{\infty}\frac{d^3p}{2\pi^3}\frac{\vp+\epsilon}{T}f_{o}}
% \end{equation}
% \begin{equation}
% 	\Rightarrow \frac{\eta}{s} \propto \tau_{\pi}
% \end{equation}

\section{Summary}
\label{sum}
Present article has attempted to build three different possible quasi-particle
models to map the QCD interaction from LQCD simulation. We have shown that 
by introducing a temperature dependent (1) degeneracy factors, (2) fugacity 
and (3) thermal width of quarks and gluons, one can build the interacting description
from non-interacting case, where those quantities had some fixed values. To fit LQCD
data of thermodynamics like entropy density, we have found the temperature dependent
parametric forms of those quantities. The degeneracy factors and fugacity are appeared
as increasing functions, whereas thermal width come as decreasing function. At infinite
temperature, they will reach their extreme limits or the non-interacting values (52, 1, 0 respectively), 
where thermodynamical quantities merge to their non-interacting values, popularly called SB limits.

After building three different forms of quasi-particle models, we have applied them
to estimate transport coefficients like shear viscosity, electrical conductivity
of QGP. Similar to the reduction of thermodynamical quantities, transport coefficients
also reduce during the transition from non-interacting to interacting picture.
The quantitative reduction of transport coefficients from different models are 
appeared to be different, although their qualitative temperature dependent curves
are quite similar. Estimating the shear viscosity to entropy density ratio,
we found that non-interacting and interacting results become same for the
degeneracy factor and fugacity based models but they are different for thermal width based model.
For latter model, we observe that interaction can have some role to reduce the value
of shear viscosity to entropy density ratio, which measure the fluid property of medium.

As a future plan, our interest to map QCD interaction in presence of magnetic field through
similar type of quasi-particle models, where we can notice the comparative roles of QCD interaction
with and without field picture on fluid property of QGP.

% At first, in this study we have estimated the dependence of degeneracy factor of the QGP 
% state to explain the crossover transition of the QGP state to hadronic phase. We have taken the 
% known degeneracy of the QGP state (Here, we have taken u,d,s quarks) then multiplied it with a 
% temperature dependent factor g(T) and tried to fit the LQCD data of the quantity $\frac{s}{T^3}$. 
% The fitting parameter g(T), which we got is an increasing function which represents the increasing 
% interactions between the quarks and gluons with decrease in temperature as degeneracy is decreasing 
% with decresing temperature for forming the hadronic matter. For explanation of this estimation, 
% we have given an analogy with the classical system of molecules i.e. changing degree of freedoms 
% with changing temperature in a system of molecules as interactions causes decrease in degree of freedom.
% 
% After that we have tried a different phenomenon to fit the LQCD data, we have taken a fugacity 
% term in the distribution functions of quarks and gluons and then tried to fit the LQCD data. 
% The Fugacity term which we have taken is temperature term. So, through fitting we have got 
% a temperature dependent fugacity term. 

{\bf Acknowledgment:} SG acknowledges to
MHRD funding via IIT Bhilai and SS acknowledges to
facilities, provided from IIT Bhilai in self-sponsor PhD scheme. 
AA, SP, RK thanks for (payment basis) hospitality from 
IIT Bhilai during his summer internship tenure (May-June, 2019). RK is supported through KVPY fellowship.

\section{Appendix}
\label{App}
\subsection{RTA method}
Here we will provide a brief framework shear viscosity ($\eta$) and electrical conductivity ($\sigma$),
whose modified estimated values due to transition from non-interacting to interacting
picture is our matter of interest in present study.
Transport coefficients like $\eta$ and $\sigma$ are basically non-equilibrium measurements
of the system by following the macroscopic relations
\bea
T^{ij}_{\eta} &=& \eta U_\eta^{ij}
\nn\\
J^i_{\sigma} &=& \sigma^{ij}E_{j}~,
\label{Mac_TJ}
\eea
where $T^{ij}$ is shear part of energy momentum tensor,
$U_{\eta}^{ij} = D^iu^j + D^ju^i +\frac{2}{3}\Delta^{ij}\partial_{\rho}u^{\rho}$
with $D^i=\partial^i -u^iu^{\sigma}\partial_{\sigma}$ and
$\Delta^{ij} = g^{ij} -u^iu^j$ is velocity gradient component, $J^i$ is electrical
current density due to electric field $E_j$.
Considering $\delta f$ deviation from equilibrium distribution function, the Eq.~(\ref{Mac_TJ})
can have microscopic form
\bea
\eta U_\eta^{ij} &=& T^{ij} =  g \int\frac{d^3p}{(2\pi)^3}\frac{p^{\mu}p^{\nu}}{E}\delta f
\nn\\
\sigma^{ij}E_{j} &=& J^i = g_e e\int\frac{d^3p}{(2\pi)^3}\frac{p^{\mu}}{E}\delta f~,
\label{Mic_TJ}
\eea
where g is degeneracy factors of medium constituents and $g_e e$ electric charge degeneracy
factors (will be discussed latter more elaborately).
Considering $\delta f$ with same tensorial decomposition,
\be
\delta f=(A_{ij}U_{\eta}^{ij} + C_iE^{i})f(1\pm f)~,
\label{del_f}
\ee
with $\pm$ stand for bosonic and fermionic medium constituents respectively and
the unknown coefficients $A_{ij}$ and $C_i$ can be found as
\bea
&&A_{ij} = \tau_{\eta} \frac{\beta p_ip_j}{E}\\
&&C_i = \tau_{\sigma} \frac{e\beta p_i}{E}
\label{AC_t}
\eea
by using the relaxation time approximation form of Boltzmann's equation.
The $\tau$ and $\tau$ are relaxation time for shear force
and electrical field respectively. Using Eqs.~(\ref{AC_t}), (\ref{del_f})
in Eq.~(\ref{Mic_TJ}), we get the final expressions of $\eta$ and $\sigma$
for gluons (boson) and quarks (fermion)~\cite{Gavin,chakrabortty,Chowdhury}:
\bea
\eta_{g,Q} &=& \frac{g_{g,Q}}{15T} \int{\frac{d^3\vp}{(2\pi)^3} \left(\frac{\vp^2}{\om}\right)^2} \tau
\: f(1\pm f)
\nn\\
\sigma_{Q} &=& \frac{e^2_Q g_e}{3T} \int{\frac{d^3\vp}{(2\pi)^3} \left(\frac{\vp}{\om}\right)^2} \tau
\: f(1 - f)~.
\label{eta_el}
\eea
Considering $g_g=16$, $g_u=24$, $g_s=12$ and $e^2_ug_e=12\times\frac{5e^2}{9}$, 
$e^2_s g_e=12\times\frac{e^2}{9}$ and $\om=\vp$, $\sqrt{\vp^2+m_u^2}$, $\sqrt{\vp^2+m_s^2}$
for $g$, $u$ and $s$, we can get total contribution of QGP as
\bea
\eta_{\rm t}&=&\eta_u+\eta_s+\eta_g
\nn\\
\sigma_{\rm t}&=&\sigma_u+\sigma_s
\eea

% where $\tau$ being the relaxation time of quarks and gluons. 
% Using the above results the final expression of the transport coefficients is written as
% 
% \bea
% \eta_{tot} && = \eta_{u,d} + \eta_s + \eta_g\nn \\
% && =  g_{u,d}\int\frac{d^3p}{(2\pi)^3}\frac{\vp^4}{E_{u,d}^2}\tau_{u,d}f_{u,d}(1-f_{u,d})\nn \\
% &&+ g_s\int\frac{d^3p}{(2\pi)^3}\frac{\vp^4}{E_s^2}\tau_sf_s(1-f_s)\nn \\
% && + g_g\int\frac{d^3p}{(2\pi)^3}\frac{\vp^4}{E_g^2}\tau_gf_g(1-f_g)
% \eea
% where $\eta_{u,d}$ , $\eta_s $ and $ \eta_g$ are the shear viscosities for $u,d$ quarks $s$ quark and gluons. $E_{u,d}$ , $E_s$ and $E_g$ are the energies 
% of $u,d$ quarks, $s$ quark and gluons given by $\sqrt{m_{u,d}^2 + \vp^2}$, $\sqrt{m_s^2 +\vp^2}$ and $\sqrt{m_g^2 + \vp^2}$. 
% 
% 
% 
% Do we have to the following expressions for $\eta$ and $\sigma$. 

\subsection{Kubo relation}
Apart from RTA method, one can obtain same expressions, given in Eq.~(\ref{eta_el}), from alternative
technique, commonly known as Kubo relation. It starts with basic definition, where any transport
coefficients ${\cal T}$ can be expressed in terms of thermal correlator of relevant operator ${\cal O}$
by a proportional relation. For ${\cal T}=\eta, \sigma$, operators are $T^{\mu\nu}_{\eta}$ and $J^\mu_{\sigma}$,
given in Eqs.~(\ref{Mac_TJ}), and their connecting
Kubo relations are~\cite{Nicola,Jeon,G_Kubo,G_el}
\bea
\eta &=& \frac{1}{20} \lim_{q_0, \vq\rightarrow 0}\frac{A_{\eta}(q_0,\vq)}{q_0}
\nn\\
\sigma &=& \frac{1}{6}\lim_{q_0, \vq\rightarrow 0} \frac{A_{\sigma}(q_0,\vq)}{q_0}~,
\label{eta_q}
\eea
where
\bea
A_{\eta}(q_0, \vq) &=& \int d^4x e^{iqx}\langle[T_{\mu\nu}(x), T^{\mu\nu}(0)] \rangle
\nn\\
A_{\sigma}(q_0,\vq) &=& \int d^4x \langle\Big[J_\mu(x), J^{\mu}(0)\Big] \rangle~.
\eea
From free Lagrangian density of quarks (fermion) and gluons (bosons), one can know 
\bea
T_{\mu\nu} &=& \Big(\Delta^{\rho}_{\mu}\Delta^{\sigma}_{\nu}-\frac{\Delta_{\mu\nu}\Delta^{\rho\sigma}}{3}\Big)
(i\bar{\psi}\gamma_{\rho}\partial_{\sigma}\psi)~{\rm for ~quark ~field}~\psi
\nn\\
&=& \Big(\Delta^{\rho}_{\mu}\Delta^{\sigma}_{\nu}-\frac{\Delta_{\mu\nu}\Delta^{\rho\sigma}}{3}\Big)
(\partial_{\rho}\phi\partial_{\sigma}\phi)~{\rm for ~gluon~field}~\phi~,
\eea
with $\Delta_{\mu\nu} = g^{\mu\nu}-u^{\mu}u^{\nu}$.
The electric current of quark field $\psi$ is
\be
J^{\mu}_Q = q_Q \bar{\psi}\gamma^{\mu}\psi~.
\ee
Using these in Eq.~(\ref{eta_q}), we can express $\eta$ and $\sigma$ in terms of field and applying
Feynman techniques, we can obtain
\bea
\eta &=& \frac{\beta}{5} \int\frac{d^3\vk}{(2\pi)^3}\frac{N_\eta}{4\omega^2\Gamma} \Big[f(1\pm f) \Big]
\nn\\
\sigma &=& \frac{\beta}{3}\int\frac{d^3\vk}{(2\pi)^3}\frac{(N_\sigma)}{4\omega^2\Gamma}[f(1-f)]
\label{eta_Kubo}
\eea
where vertex kind of factor of respective one-loop self-energy diagrams can be obtained as
\bea
N_\eta &=& g\frac{4\vk^4}{3} 
\nn\\
N_\sigma &=& g_e4e_Q^2\vk^2
\label{N_k}
\eea
and thermal width $\Gamma=1/\tau$ has been introduced in propagators of respective diagrams to
cure their divergence near zero momentum. After putting Eq.~(\ref{N_k}) in Eq.~(\ref{eta_Kubo}),
one can return to Eq.~(\ref{eta_el}). So we can get same final expressions of $\eta$, $\sigma$ 
from either kinetic theory of RTA method or Kubo-type (one-loop) diagrammatic method.

\end{document}